# Amplification of Interstellar Magnetic Fields by Supernova-Driven Turbulence

By

Dinshaw S. Balsara[1], Jongsoo Kim[2], Mordecai-Mark Mac Low[3] and Grant J. Mathews[1]

(dbalsara@nd.edu, jskim@kao.re.kr, mordecai@amnh.org, gmathews@nd.edu)

[1]Center for Astrophysics, Department of Physics, University of Notre Dame,

[2]Korea Astronomy Observatory

[3]Department of Astrophysics, American Museum of Natural History

**Abstract**

Several lines of evidence suggest that magnetic fields grow rapidly in protogalactic and galactic environments. However, mean field dynamo theory has always suggested that the magnetic fields grow rather slowly, taking of order a Hubble time to reach observed values. The theoretical difficulties only become worse when the system has a high magnetic Reynold's number, as is the case for galactic and protogalactic environments. The discrepancy can be reconciled if fast processes for amplifying magnetic field could operate. Following Balsara, Benjamin & Cox (2001), we show that an interstellar medium that is dominated by realistic energy input from supernova explosions will naturally become a strongly turbulent medium with large positive and negative values of the kinetic helicity. Even though the medium is driven by compressible motions, the kinetic energy in this high Mach number flow is mainly concentrated in solenoidal rather than compressible motions. These results stem from the interaction of strong shocks with each other and with the interstellar turbulence they self-consistently generate in our model. Moreover, this interaction also generates large kinetic helicities of either sign. The turbulent flow that we model has two other characteristics of a fast dynamo: magnetic energy growth independent of scale, and with a growth time that is comparable to the eddy turn-over time. This linear phase of growth permits the field to grow rapidly until the magnetic energy reaches about 1% of the kinetic energy. At that stage, other astrophysical processes for producing magnetic fields can take over. Energetics, power spectra, statistics and structures of the turbulent flow are studied here. Shock-turbulence interaction is shown to be a very general mechanism for helicity generation and magnetic field amplification with applicability to damped Ly-$\alpha$ systems, protogalaxies, the Galaxy, starburst galaxies, the inter-cluster medium and molecular clouds.

## I) Introduction

Ever since the early observation of finite rotation measures in high-redshift quasars by Kronberg & Perry (1982), Perry, Watson & Kronberg (1993) and Kronberg (1995) there has been considerable interest in the formation of magnetic fields in such environments. This interest has increased as a result of the observation of fields of a few µG in damped Ly-α systems at *z=2* by Wolfe, Lanzetta & Oren (1992). The result is doubly interesting. First, it shows that high-redshift systems that may or may not have a well-established differential rotation still manage to form magnetic fields with field strengths comparable to our present galactic field strength (see Rand & Kulkarni 1989; Beck et al. 1996). Second, it shows that the field reaches this strength extremely fast compared to the several gigayear e-folding times estimated for mean field dynamos (e.g. Ferrière & Schmidt 2000).

A study of the magnetic field in our own galaxy yields further insights. Rand & Kulkarni (1989) studied the magnetic field within a few kiloparsecs of the Solar neighborhood. They found that the mean magnetic field had a magnitude of 1.6 µG while the fluctuating component of the magnetic field was estimated to have a strength of about 5 µG. Beck et al. (1996) carried out similar observations of the Galaxy and nearby galaxies. They too found that the fluctuating component of the magnetic field had a strength that was comparable to the mean field. These observations, however, only probe the fluctuating field on a very limited range of length scales. Nevertheless, the presence of a fluctuating component may be explained as resulting from turbulent motions in the interstellar medium (ISM). Observations by Spangler (1999), Haverkorn, Katgert & de Bruyn (2000) and Gaensler et al (2001) probe the structure of the ISM on a much larger range of length scales. The latter observations further support the idea that the ISM might indeed be turbulent, though they do not directly probe the structure of the magnetic field. Minter and Spangler (1996) observed rotation and emission measures for a small section of the sky to deduce the spectrum of magnetic field fluctuations on scales smaller than 100 pc, finding evidence for a turbulent component in the magnetic field. Fosalba et al (2002) used starlight polarization to come to a similar conclusion. Han, Ferriere and Manchester (2004) used pulsar dispersion and rotation measurements to deduce the energy spectrum of the interstellar magnetic fields on scales spanning 0.5 – 15 Kpc. The data suggests that the turbulence is three-dimensional on smaller scales but becomes two-dimensional on larger scales. There is no current observational evidence for the structure of magnetic fields on a range of proto-galactic length scales. However, there is adequately strong evidence for vigorous star formation activity leading to turbulence in such environments, which suggests that magnetic field fluctuations would develop on a range of length scales in those systems too. It is, therefore, interesting to study the formation and evolution of magnetic fields in such turbulent environments.

The papers by Field, Goldsmith & Habing (1969), Cox & Smith (1974) and McKee & Ostriker (1977), when taken together, have shown that the three phase structure of the interstellar medium is primarily sustained by energetic input from supernova (SN) explosions and winds from OB stars. This is because an accounting of the energetic input into the ISM shows that SNe and winds, in that order, are the dominant sources of mechanical input in the turbulent ISM (see reviews by Mac Low & Klessen 2004, Elmegreen and Scalo 2004 and Scalo and Elmegreen 2004). Balsara, Benjamin & Cox (2001; hereafter BBC) have studied the interaction of three dimensional SN remnants with a turbulent, magnetized ISM. Korpi et al. (1999) carried out low resolution simulations of SN-driven, magnetized interstellar turbulence. Kim, Balsara & Mac Low (2001) studied the energetics, structure and spectra of such a turbulent medium with substantially higher resolution simulations. Balsara & Kim (2004) made a detailed study of the role of good numerical methods in enabling a faithful representation of magnetic field amplification in such SN-driven turbulence. Mac Low et al. (2004) have studied the thermodynamic nature of interstellar turbulence with SNe being the dominant energy input mechanism. The simulation work cited above clearly indicates that the turbulence is supersonic and that compressibility effects produce a range of densities. It is the purpose of this paper to make a detailed study of the magnetic field amplification in such a turbulent, compressible, magnetized ISM with energy input being provided by SN explosions. Understanding this problem may yield insights into magnetic field generation in primordial galaxies, starburst galaxies, our Galaxy and perhaps even our Galactic center.

Ruzmaikin, Shukurov & Sokoloff (1988) have used mean field dynamo theory to study the problem of magnetic field generation in the Galaxy. Mean field dynamo theory is based on the early work of Parker (1971) and Steenbeck, Krause & Radler (1966) who formulated the problem for incompressible magnetohydrodynamics (MHD) by assuming that one could effect a split in length scales. On the small scales, the magnetic fields and helical fluid motions are presumed to be driven by several forms of gentle convective motion. These gentle helical motions, along with the mean shear in the Galactic rotation curve, produce growth of the mean field on large scales. To sustain a persistent dynamo it is, therefore, important to find robust mechanisms for helicity generation in the Galaxy. Ferrière (1993ab, 1995, 1998) showed that SN explosions and superbubbles in the sheared and stratified Galactic ISM could provide one possible mechanism for helicity generation. This mechanism for generating helicity was used in conjunction with the theory for the $\alpha$-$\omega$ dynamo by Ferrière & Schmitt (2000) to predict growth times for the mean Galactic field. The time of growth for the large-scale magnetic field was found to be about 1.8 Gyr for most of their models, suggesting that the Galactic magnetic field has presently reached a value that is close to saturation. The large times of growth indicate that the Galactic magnetic field has undergone no more than a few e-foldings over a Hubble time. To reach its current strength in the Galaxy, mean field dynamo theory, therefore, indicates that the protogalaxy should have started with a magnetic field that was ~ $1.5 \times 10^{-9}$ G. However, simulations of the Biermann (1950) battery effect and its ability to produce strong seed fields in clusters and protogalactic environments show that the protogalactic seed field should be much smaller, see Kulsrud et al. (1997).

The simulations of Kulsrud et al. (1997) find that the protogalactic fields were ~ $10^{-21}$ G. This suggests that it might be useful to look for field amplification mechanisms that go beyond the limiting assumptions of mean field dynamo theory. For example, mean field dynamo theory is predicated on slow, almost incompressible motions that take place in a non-clumpy, almost smooth plasma. Observations suggest that the ISM has supersonic motions that produce strong local compressions. Furthermore, the highly compressible equation of state of the ISM gives it a strong tendency to form clumpy clouds of cold gas, each of which has its own internal turbulence. A thorough study of blast wave interactions and their role in generating kinetic helicity and amplifying magnetic fields in strongly compressible, clumpy, radiatively cooled and heated, turbulent media has never been undertaken. We begin such a study here.

Kulsrud & Anderson (1992; hereafter KA) pointed out that there is a further problem with mean field dynamo theory when it is applied to galactic/protogalactic environments. KA's calculations are based on Kraichnan & Nagarajan's (1967) LDIA approximation for incompressible MHD turbulence. KA point out that the magnetic Reynolds number ($R_m$) of our Galaxy is very large, $R_m \sim 10^{16}$. <u>KA conclude that the large value of $R_m$ implies that mean field dynamo theory is not be applicable to our Galaxy.</u> The high value for $R_m$ causes turbulent velocity fluctuations and turbulent magnetic field fluctuations to develop on small scales. These small-scale fluctuations in the magnetic field grow rapidly and the calculations of KA suggest that the small-scale magnetic energy should grow rapidly to the point where it is in equipartition with the kinetic energy. This, in turn, was claimed to quench the turbulence, thereby arresting the growth of large scale magnetic fields. One of the suggestions that comes out of the work of KA is that it is very important to study the growth of magnetic fields in systems where there is a considerable separation of scales between the large scales and the small scales. Thus, simulating the growth of magnetic fields in computational domains that are large enough to permit a separation of scales between the large scales and the small scales is a problem of great interest. KA also ignore the fact that interstellar turbulence is driven by strong SN shocks. These shocks are strong enough to overpower the small-scale saturation. While KA and Howard & Kulsrud (1997) have shown the importance of a clumpy medium, they have not paid much attention to the role of strong shocks when they interact with such a clumpy ISM. We show the importance of such interactions in the next paragraph.

As seen from the work of Ferrière (1993ab, 1995, 1998), the existence of robust mechanisms for generating kinetic helicity is of great interest in magnetic field amplification. McKenzie & Westphal (1968) were the first to realize the importance of shocks and their ability to amplify turbulence when they interact with turbulence. McKee & Zweibel (1995) came to a similar realization in the astrophysical literature. Samtaney & Zabusky (1994) used shock polar analysis to calculate scaling relations for vorticity generation when shocks interact with density inhomogeneities in two dimensions. They found that increasing shock Mach numbers or increasing density contrasts enhance vorticity generation at shocks. While kinetic helicity cannot be generated by two-dimensional flows, a robust mechanism for vorticity-generation in two dimensions will produce helicity when turbulent motions in the third dimension are allowed. While the amplification of turbulence that is demonstrated in McKenzie & Westphal (1968) is based on a linear analysis, the demonstration of vorticity generation by Samtaney & Zabusky (1994) is an inherently non-linear calculation. As a result, the generation of helicity when shocks interact with turbulence should be a robust conclusion even in the non-linear regime. BBC showed that the interaction of SN shocks with interstellar turbulence amplifies post-shock turbulence and is also a source of helicity-generation. They showed that strong fluctuations in kinetic helicity of either sign are produced when the turbulence is homogeneous and isotropic. Supernova remnants and superbubbles can interact with large-scale density contrasts in the ISM resulting in a break-out of the remnant's cavity (e.g. Magnier et al. 1996, Williams et al. 1999). This is different from the conventional description for the terminal evolution of a remnant by Cox and Smith (1974) which postulates that the remnant begins to merge gradually with the ambient medium when its expansion speed equals the ambient medium's sound speed. However, various instability mechanisms, such as the Vishniac overstability (Vishniac 1983), the Rayleigh-Taylor instability during breakout (e.g. Mac Low and McCray 1988) and the gravitational instability (e.g. McCray and Kafatos 1987), might be other mechanisms for helicity production in supersonic, compressible, clumpy, interstellar. One must, therefore, pay attention to those other sources of helicity production.

There is also a substantial amount of recent interest in studying magnetic field generation in helically driven turbulence. Such turbulence would constitute an $\alpha^2$ dynamo. Such a study of the turbulent dynamo with scale separation was reported by Zeinecke, Politano & Pouquet (1998; hereafter ZPP) for the case of incompressible MHD and was extended to compressible MHD using even larger computational domains by Balsara (2000). Both ZPP and Balsara (2000) studied a turbulent MHD system that was driven with motions that had the maximal amount of kinetic helicity with a single sign. They found that the initial magnetic field energy grew exponentially with a constant growth rate. They called this epoch the regime of linear growth. As the field strength grew, they found that the rate of growth of the magnetic energy was diminished when the magnetic energy reached a strength of ~2% of the kinetic energy. However, the growth of the field persisted at a slower rate so that the magnetic field underwent what they called a quasi-linear phase of growth. This happened despite the fact that the small scale magnetic field strength was less than an order of magnitude away from equipartition with the velocity field. Both ZPP and Balsara (2000) were able to show that systems existed that could approach equipartition between the magnetic energy and the kinetic energy and yet show large scale field growth as long as the helical driving was persistent. While the ZPP and Balsara (2000) simulations were maximally helical, Maron & Blackman (2002) have done incompressible simulations with fractionally helical driving and found that when the fraction of helical driving exceeds a certain value, large scale field growth takes place. Galloway and Proctor (1992) studied the evolution of a kinematic dynamo in a flow with Lagrangian chaos and zero mean helicity and found evidence for magnetic field amplification. Since Maron & Blackman (2002) did not analyze the Lagrangian chaos in their streamlines it is difficult to connect the two results. All of the simulations cited in this paragraph are based on idealized forms of driving that were put in by hand. None of them was based on SN-driven turbulence of the sort that occurs naturally in the ISM. Thus it is of interest to examine what SN-driven turbulence might do.

The driven turbulence studied by ZPP and Balsara (2000) was motivated by a class of theories known as fast dynamo theories. Fast dynamo theories specifically apply to the high $R_m$ regime and were first proposed by Vainshtein & Zeldovich (1972).

Childress & Gilbert (1995) have reviewed such theories. They find that the magnetic field can grow very rapidly in certain high $R_m$ systems that have a net helicity. The amplification of the magnetic field takes place via a "stretch, twist and fold" (STF) mechanism. Childress & Gilbert (1995) also realized the limitations of such a forcing. Two persistent questions come through in Childress & Gilbert (1995). They are: a) Are there physical systems that can produce large helicity fluctuations of either sign so that the overall helicity of the system is zero? b) Can one have fast magnetic field growth in such systems? The simulations that are presented in the current paper show both that the ISM is a physical system that might naturally produce strong helicity fluctuations of either sign. In a subsequent paper, Balsara, Kim and Mathews (2004), we will address the issue of rapid flux growth and its relation to dynamical chaos.

In Section 2 we describe the physical system and provide numerical details. In Section 3 we study the energetics associated with magnetic field growth in such systems. In Section 4 we study spectra of magnetic and kinetic energy. In Section 5 we study turbulent structures and statistics that result in the system being studied. In Section 6 we offer a discussion and some conclusions.

**2) Description of the Numerical Model**

In the first part of this section we describe the basic equations and the numerics used. In the second part we describe the simulations.

**2.a) Basic Equations**

We solve the three dimensional MHD equations in conservation form with heating and cooling adjusted for the ISM as:

$$\frac{\partial}{\partial t}\begin{pmatrix} \rho \\ \rho v_x \\ \rho v_y \\ \rho v_z \\ \varepsilon \\ B_x \\ B_y \\ B_z \end{pmatrix} + \frac{\partial}{\partial x}\begin{pmatrix} \rho v_x \\ \rho v_x^2 + P + \mathbf{B}^2/8\pi - B_x^2/4\pi \\ \rho v_x v_y - B_x B_y/4\pi \\ \rho v_x v_z - B_x B_z/4\pi \\ (\varepsilon+P+\mathbf{B}^2/8\pi)v_x - B_x(\mathbf{v}\cdot\mathbf{B})/4\pi \\ 0 \\ (v_x B_y - v_y B_x) \\ -(v_z B_x - v_x B_z) \end{pmatrix}$$

$$+ \frac{\partial}{\partial y}\begin{pmatrix} \rho v_y \\ \rho v_x v_y - B_x B_y/4\pi \\ \rho v_y^2 + P + \mathbf{B}^2/8\pi - B_y^2/4\pi \\ \rho v_y v_z - B_y B_z/4\pi \\ (\varepsilon+P+\mathbf{B}^2/8\pi)v_y - B_y(\mathbf{v}\cdot\mathbf{B})/4\pi \\ -(v_x B_y - v_y B_x) \\ 0 \\ (v_y B_z - v_z B_y) \end{pmatrix} + \frac{\partial}{\partial z}\begin{pmatrix} \rho v_z \\ \rho v_x v_z - B_x B_z/4\pi \\ \rho v_y v_z - B_y B_z/4\pi \\ \rho v_z^2 + P + \mathbf{B}^2/8\pi - B_z^2/4\pi \\ (\varepsilon+P+\mathbf{B}^2/8\pi)v_z - B_z(\mathbf{v}\cdot\mathbf{B})/4\pi \\ (v_z B_x - v_x B_z) \\ -(v_y B_z - v_z B_y) \\ 0 \end{pmatrix}$$

$$= \begin{pmatrix} 0 \\ 0 \\ 0 \\ 0 \\ -\rho^2 \Lambda(T) + \rho\,\Gamma_0 \\ 0 \\ 0 \\ 0 \end{pmatrix}$$

where $\varepsilon = \rho v^2/2 + P/(\gamma-1) + \mathbf{B}^2/8\pi$ is the total energy of the plasma. In all our simulations the ratio of specific heats, $\gamma$, was set to 5/3.

The heating is representative of photoelectric and cosmic ray heating. The cooling is obtained from the work of Raymond & Smith (1977). As a result, the cooling function used in this work does not incorporate some of the molecular cooling terms, and especially the heating effects of PAHs, as shown in Wolfire et al. (1995, 2003). As a result, the current round of simulations have the requisite physics to produce the hot and the warm phases of the ISM but not the cold phase. Inclusion of the cold phase, along with all the extra physics and numerics that is needed for its inclusion, will be the topic of subsequent papers. Because we have chosen solar metallicities in this work, the simulations are directly applicable to the Galactic ISM. However, it is useful to point out that Mushotsky and Loewenstein (1997) have observed several clusters and found them to have a metallicity that is ~0.3 of our solar value. Likewise, Churchill and Le Brun (1998) and Ledoux, Srianand, and Petitjean (2002) find metallicities in damped Lyman α systems that are close to solar values. Thus, this work will also have applicability to these other systems.

The equations are solved on a three dimensional Cartesian mesh using the TVD methods described in Balsara & Spicer (1999a,b), Balsara (2004) and Balsara & Kim (2004). Specifically, we use the fast TVD algorithm drawn from the RIEMANN code that was described in Balsara (2004) and calibrated for the present application in Balsara & Kim (2004). The solution methods are all based on higher order Godunov schemes that are known to produce effective magnetic Prandtl numbers of unity. While that may be different from the Galactic value, the methods are the only known methods that will robustly integrate the physics of strong shocks that needs to be represented for this scientific problem.

**2.b) Description of Simulations**

The simulations evolve small cubical patches of the ISM that are 200 pc on an edge. This size is chosen because a distance of 100 pc is smaller than the scale height of the ISM. Hence, the simulations represent a patch of the ISM in the midplane of the Galaxy. Balancing heating with cooling at a specified density and temperature enables us to set $\Gamma_0$, the assumed rate of photoelectric and cosmic ray heating. Thus, the mean density and temperature are two of the parameters that specify these simulations. We start with a very small, uniform, seed magnetic field, which also constitutes one of the parameters that specifies the simulations. The SN rate is another determining parameter in these simulations. In this work we adopt a fiducial supernova rate in our Galaxy of two SN explosions per century. Thus a Galactic SN rate corresponds to one supernova exploding every 1.26 Myr in our computational domain. For our purposes there is no need to distinguish between Type I and Type II SNe. Thus, the SN rate is specified as a multiple of this Galactic SN rate. As a first approximation, the SNe explode in random locations in the computational domain. For our present computational purposes, a SN explosion imparts $10^{51}$ ergs of thermal energy within a radius of 5 parsecs of the randomly selected explosion site. That energy pulse then evolves in time. The SN rate determines the interval of time between successive SNe. Extensive code tests were done to ensure that isolated SN remnants that were initialized with a radius of 5 parsecs evolved spherically in a uniform, unmagnetized medium. To retain physical consistency across simulations, the same initial radius of 5 parsecs was used for the remnants on all the meshes that were used in this work.

The simulations tend to be very long-running even on the fastest parallel supercomputers, with typical run times that exceed $10^5$ cpu-hours. This is so because, for a field growth experiment one needs to run the simulations for several eddy turn-over times, resulting in almost a thousand SNe going off in the course of a run. The short timesteps needed to follow each individual SN remnant along with the large number of explosions makes these simulations very challenging. We also wish to have results that are close enough to the converged results of an infinitely fine mesh. As a result, we present results from two identical simulations that have resolutions of $128^3$ and $256^3$ zones. If the two simulations show roughly similar systematic behavior, we can conclude that the answers we obtain from the higher resolution simulation are close enough to the converged answers. Initially, we carried out three simulations at $128^3$ zone resolution with eight, twelve and forty times the galactic SN rate. The most interesting of these simulations was then repeated on a $256^3$ zone mesh. That simulation had a density of 1 amu / cm$^3$ , a mean temperature of 10,000 K, an initial magnetic field energy that was $2\times10^{-6}$ times smaller than the thermal energy and a SN rate that was 8 times larger than our adopted Galactic rate. The higher SN rate was necessary for producing appreciable field growth in a sufficiently short simulational time. Lower SN rates will also produce field growth, albeit at a somewhat slower rate. Moreover, such high SN rates are also consistent with those expected to occur in starburst and protogalactic environments.

**3) Study of the Energetic Aspects of the Turbulence**

**3.a) Energy Growth Rates**

Figure 1 shows the growth of magnetic energy <u>and rms density fluctuations</u> for runs with $128^3$ and $256^3$ zones for the first 40 Myr of simulation time. Both runs show a very rapid initial increase in the magnetic energy during the first 5 Myr. This initial growth does not represent the growth of magnetic energy in steady state turbulence. Instead, the first 5 Myr in both simulations correspond to the time it takes for the SN-driven forcing to produce well-developed turbulence, <u>as shown by the saturation of the rms density fluctuations</u>. The first 5 Myrs, therefore, correspond to the time it takes for every parcel of gas in the simulation to be fully processed in at least one SN. A higher rate of SNe results in a faster onset of fully-developed turbulence, a lower rate of SNe results in a slower onset of fully-developed turbulence. To verify that we have fully-developed turbulence in the computational domain, we compared the velocity spectrum at 5 Myr to the velocity spectrum at later times and found that the spectral shape remains unchanged. <u>The growth of magnetic energy after 5 Myr, despite the saturation of the rms density fluctuations, allows us to safely assert that the magnetic field growth does occur in steady-state turbulence and is not due to rms density fluctuations that increase with time.</u>

After the first 5 Myrs the field energy continues to grow almost monotonically, increasing by over two orders of magnitude in the course of the simulation. The temporal evolution of magnetic energy, shown in Figure 1, displays jitters over very short time intervals. To eliminate the influence of these jitters in quantifying the growth time of the magnetic energy, we used a sliding window of 10 Myr to measure the best-fit value of the growth time. The growth time is the time required by the field to increase by one e-folding. In Figure 2 we plot growth time as a function of simulation time for both runs.

We see from Figure 2 that the magnetic field in the $256^3$ zone run has growth time fluctuating from 4 to 16 Myr in the time interval from 10 to 40 Myr. We also see that the growth time is not a constant, but rather fluctuates in a broad range as a function of time. The fluctuations appear to be a consequence of the random locations of the SNe.

It is very instructive to compare the $128^3$ and $256^3$ zone simulations. From Figure 1 we see that the $128^3$ zone simulation has a slightly lower value of the magnetic energy than the $256^3$ zone simulation. This owes to the fact that the $256^3$ zone simulation has many more zones than the $128^3$ zone simulation. Hence, there is a lot more small-scale magnetic structure that can be represented in the $256^3$ zone simulation which is not present in the $128^3$ zone simulation. The $256^3$ zone simulation, therefore, seems to contain more magnetic energy than the $128^3$ zone simulation. This is simply a manifestation of the fact that neither of the two simulations has enough resolution to permit a long inertial range to form. Nevertheless, the curves of growth do track each other, indicating that there is a rough agreement between the two simulations. Likewise, both simulations show growth of magnetic energy over several orders of magnitude, indicating that the growth of magnetic energy in our simulated ISM is a robust conclusion. From Figure 2 we also see that the growth times for the two simulations track each other for a good fraction of the interval. The only exception is the interval between 25 Myr and 36 Myr. On detailed examination of the placement of successive SN explosions we found that the random number generator produced SN explosions that were contiguous to each other at 25 Myr. On the larger mesh we found that the explosions still have sufficient distance between each other that one explosion did not go off within the cavity of the previous explosion. On the smaller mesh, that was not the case, causing unusually high rates of heating, and a subsequent slow down in the growth of magnetic field. This shows us two things: First, it shows that there are numerically induced differences that can be triggered in this problem which require careful subsequent interpretation. Second, it shows us that SN explosions going off within the cavities of previous remnants are almost inevitable, especially when the SN rate is high enough. When such episodes occur, one might experience a local decrease in the growth rate of the magnetic field. This point will be expanded upon later in Section 3.d.

Figure 1 shows that the energy in the magnetic field energy increases by over two orders of magnitude in the course of the simulation but does not reach equipartition. In fact, by the end of the simulation the field energy is still only 1% of its equipartition value. We do, however, see that the field is still growing at a robust rate towards the end of the simulation.. Using more idealized simulations, ZPP and Balsara (2000) both found that once the magnetic energy reaches 2% of the kinetic energy the growth rate of the magnetic energy slows down markedly. They called this the quasilinear phase of evolution. The field underwent a protracted phase of quasilinear evolution (with a much reduced growth rate) before the magnetic energy reached rough equipartition with the kinetic energy. For that idealized problem it was very important to clearly demonstrate that turbulent flows could indeed amplify the field energy to equipartition. In our present simulations we too seem to have reached a stage where the magnetic energy has reached about 1% of the kinetic energy by the end of 40 Myr. In this problem it is not important to demonstrate that equipartition is achieved. This is because once the field strength has been raised to appreciable levels in a protogalaxy, slower processes, such as the $\alpha-\omega$ dynamo, can take over and increase the large scale field even further. We also factor in the practical consideration that our higher resolution simulation took over four months to run on a medium-sized PC cluster. As a result, we will leave the issue of approach to equipartition and the role of other astrophysical processes in producing further field growth for future work.

**3.b) Similarities to a Fast Dynamo**

In Balsara, Kim and Mathews (2004) we will analyze the Lyapunov exponents associated with streamlines, thereby demonstrating the Lagrangian chaos in the SN-driven turbulence. We will also make a detailed connection between the rapid growth of magnetic flux and various indicators of Lagrangian chaos in the flow. ZPP and Balsara (2000) argue that certain well-designed energy considerations provide evidence of a fast dynamo. Since this section is devoted to energy considerations, we repeat the same for our own simulations below.

First is a growth time for magnetic energy comparable to the eddy turn-over time. The density-averaged rms velocity in our simulations $v_{rms} \sim 12$ km s$^{-1}$. Using a median SN remnant radius $R \sim 30$ pc, we find an effective eddy turn-over time $R/v_{rms} \sim 2.5$ Myr, of the same order as the growth times shown in Figure 2. ZPP mention that the existence of a magnetic energy growth rate that is comparable to the eddy turn-over time is one of the circumstantial pieces of evidence that a fast dynamo is operating. The growth times we find in this paper are much shorter than the $\sim 1.8$ Gyr growth times that were estimated by Ferrière & Schmidt (2000) for the $\alpha-\omega$ dynamo driven by Galactic superbubbles.

A second feature is a magnetic energy growth rate that is independent of scale, which does not arise in the classical $\alpha^2$ dynamo. ZPP and Balsara (2000) also used spectral analysis to show that the magnetic energy on long scales and short scales grew at the same rate in the linear regime. They showed that a scale-independent growth rate provides another piece of evidence indicating that a fast dynamo-like behavior was being observed. Owing to the simplicity of their forcing term, ZPP and Balsara (2000) could clearly identify a length scale on which they forced the turbulent flow. They identified the long scales as scales that exceeded their forcing scale, and short scales as the rest. In the present problem, the size of a SN remnant takes the place of a forcing length scale. Because SN remnants occupy a range of length scales, one cannot identify a single forcing scale in the present problem. We, therefore, identify the radius of a typical remnant , 30 pc, as the scale at which we distinguish between large and small scales. A length of 30 pc corresponds to one-seventh of our computational domain. Figure 3 shows the growth of magnetic energy on all scales that are larger than 30 pc (solid line) and on all scales that are smaller than 30 pc (dashed line). <u>From 5 to 20 Myr we do see that the short scales grow somewhat faster than the longer scales, owing to the fact that the direct cascade of magnetic energy has shorter eddy turnover times and can establish itself much more rapidly. However, past 20 Myrs we see that the two curves do indeed grow at the same rate for an appreciable length of time.</u>

While the ad hoc split into small and large length scales in this problem may seem arbitrary, it is possible to analyze the rate of growth for several of the individual spectral modes in the problem. This is done in Figure 4 which shows the temporal evolution of the k = 3 and 5 spectral modes. These represent length scales that are decidedly larger than the size of a SN remnant. We also plot out the temporal evolution of the k = 9 and 11 spectral modes which represent length scales that are decidedly smaller than the size of a SN remnant. It is easily seen that past 20 Myrs the large scale modes and the small scale modes grow at the same rate.

**3.c) Kinetic and Magnetic Helicity**

While the mean kinetic helicity in this problem remains zero, it is interesting to ask how it evolves? We also realize that the magnetic energy grows with time in this problem. The strong connection between the inverse cascade of magnetic helicity and the problem of magnetic field amplification was first pointed out by Pouquet, Frisch and Leorat (1976). Balsara and Pouquet (1999) have also carried out simulations of the inverse cascade of magnetic helicity for the compressible case. As a result, the evolution of magnetic helicity is also a topic of interest. Figure 5 plots the evolution of the rms fluctuations in the kinetic and magnetic helicities on the same plot as a function of time. The kinetic helicity is given by $\mathbf{v} \cdot \nabla \times \mathbf{v}$. The magnetic helicity is given by $\mathbf{A} \cdot \mathbf{B}$ where $\mathbf{B} = \nabla \times \mathbf{A}$. The kinetic helicity shows a very spiky behavior, where each spike corresponds to a SN explosion, supporting the suggestion that the interaction of the remnants with the turbulent ISM is a strong source of kinetic helicity. We also see that the rms fluctuation in the kinetic helicity does not grow, unlike the rms fluctuation in magnetic helicity, which does. This is consistent with the fact that the magnetic energy grows with time while the kinetic energy does not.

**3.d) Effect of Supernova Explosion Rate**

The simulations that we have discussed so far all had a SN rate of eight times Galactic. Will this mechanism persist at much higher SN rates? Many real-world astrophysical effects could change the answer that we attempt to give here. For example, stratification of the disks in real galaxies will result in the hot phase of the ISM blowing out via bubbles and chimneys, as observed in NGC 891 (Rand, Kulkarni, & Hester 1991). However, over the course of our simulations, we did explore the conditions that are needed for the present mechanism to operate via a brief parameter study. Figure 6(*a-c*) shows the evolution of thermal, kinetic and total energy as a function of time in runs with SN rates of eight, twelve, and forty times the Galactic value, while Figure 6*(d)* shows the evolution of magnetic energy in the last run. When the SN rate exceeds a critical rate, we expect SN explosions to occur frequently within the cavities of previous remnants. When that happens, we expect the magnetic energy growth to be quenched. We can only bracket the critical rate with the simulations in hand. In Figures 6*(a)* and *(b)* we see that the runs reach thermal equilibrium where energy input from SNs balances energy loss from radiative cooling. In those cases we find robust growth of magnetic field. <u>Note though that the magnetic energy is substantially smaller than the kinetic and thermal energies all through the evolution of these simulations. Thus one should not expect to see a secular growth in total energy as a result of growth in magnetic energy.</u> In Figure 6*(c)* a thermal runaway sets in as all the gas is heated to temperatures where radiative cooling becomes inefficient. <u>The thermal runaway is initiated at 2 Myr but it takes till about 8 Myr before most of the gas is converted to million degree gas, making the runaway pervasive.</u> Figure 6*(d)* shows that the magnetic energy in this case does grow initially, but is quenched once the thermal runaway becomes pervasive. The critical SN rate thus lies somewhere between twelve and forty times the Galactic value for our chosen initial density and temperature.

That such a quenching should exist past some critical SN rate is easy to understand on physical grounds. The growth of field is based on helicity production when strong SN shocks interact with a predominantly warm ISM, as shown by BBC. However, after thermal runaway, most of the ISM ceases to be warm. In fact, most of it is filled with million degree hot gas. When a SN explosion occurs in such a hot medium the shock quickly becomes weak and so stops producing helicity via its interaction with ambient turbulence; therefore magnetic field amplification ceases.

## 4) Power Spectra of the Turbulence

Having studied the bulk energetics of the system, we now turn our attention to the spectral domain. Figure 7 shows the power spectra for kinetic and magnetic energies and the density at several different times in the $256^3$ zone run. The kinetic energy and the density saturate while the magnetic energy keeps growing. <u>The saturation of the density spectra on all scales provides further evidence that the magnetic field growth past 5 Myr is not a consequence of time-dependent rms fluctuations in the density.</u> We see, however, the magnetic energies have not reached equipartition even on small scales by the end of the simulation. This shows that the present simulations do not explore saturation of turbulent field amplification. The temporal evolution of the magnetic energy closely mirrors the energetic evolution of the small and large scale magnetic fields that was shown in Figures 3 and 4.

A fair bit of work has been done to study turbulence associated with strong shocks. In our calculation we drive the turbulence with very strong shocks indeed. It is, therefore, of some interest to analyze the kinetic energy spectrum shown in Figure 7. In Figure 8, we decompose the spectrum into its compressible and solenoidal parts. Different epochs are shown with different colors. The solid lines correspond to the solenoidal part of the velocity spectrum, the dotted lines correspond to the compressible part of the velocity spectrum. We find the surprising result that at most length scales, the solenoidal part of the velocity spectrum overwhelms the compressive part of the velocity spectrum by almost an order of magnitude. This is <u>quite an interesting</u> result, given that the computations were forced with very strong shocks. <u>Similar results for strongly forced turbulence have been reported in Passot and Pouquet (1987), Balsara and Pouquet (1999), Porter et al (1999), Boldyrev (2002) and Vestuto et al (2003).</u> It is also significant that Fig. 6*(a)* shows that the turbulence has a Mach number that is in excess of unity. Despite this high Mach number, Figure 8 shows that the turbulence can sustain strong vortical motions. Figure 8 also provides a spectral confirmation of the demonstration by BBC that strong shocks interacting with a clumpy and turbulent ISM can be a strong source of kinetic helicity fluctuations. We put three line segments with slopes of –1.5, –1.67 and –2 in Figure 8 which correspond to the spectral indices of Kolmogorov, Iroshnikov and Kraichnan and Burgers turbulence respectively. Because of the dominance of SN shocks, both the solenoidal and compressible spectra have inertial ranges with spectral indices that are close to two.

Figure 9*(a)* shows the absolute value of the magnetic helicity spectrum at various times in the simulation. Figure 9*(b)* shows similar spectra for the kinetic helicity. We see quite clearly that, despite their jaggedness, the magnetic helicity spectra show a secular increase with increasing time, consistent with the growth in magnetic energy. Figure 9*(b)* shows no such secular increase. These results are consistent with Figure 5 which shows that the rms fluctuations in the kinetic helicity do not show secular growth as a function of time while the rms fluctuations in the magnetic helicity do indeed show secular growth as a function of time.

**5) Structures and Statistics of the Turbulence**

Figure 10 shows images of the log of thermal and magnetic pressure, and of the log of kinetic helicity and the magnetic helicity on a slice plane that passes through a recent remnant (the one at the bottom of the panel). This particular remnant exploded in a rather low density environment. The kinetic helicity is colored so that large negative values of the helicity appear dark blue while large positive values of the helicity appear deep red. The figures correspond to a time of 20 Myr by which time the simulated ISM is strongly turbulent. We observe that the remnant is not spherical and, in this strongly turbulent environment, the remnant is even more aspherical than the remnants simulated in BBC. The turbulence that gets established in the simulations reported here is much stronger than the turbulence in BBC. As a result, we see substantially more aspherical evolution in the SN remnants simulated here than in BBC. The images for the thermal pressure and kinetic helicity in Figure 10 when taken together clearly show that the largest fluctuations in the kinetic helicity are somewhat correlated with the remnant. It is, however, noteworthy that there are many other locations in the simulation that are disjoint from the remnant which also show quite large fluctuations in kinetic helicity. Figure 10 shows that the magnetic field does get compressed at the remnant's boundaries. However, large scale strongly magnetized structures are visible in all parts of the domain. This shows us in a very graphical fashion that growth of magnetic field structures has taken place all over the computational domain. The magnetic helicity in Figure 10 shows substantial correlation with the magnetic energy, as expected. A similar result was also found in BBC.

## 6) Discussion and Conclusions

### 6.a) Discussion

The mechanisms for field generation demonstrated here are potentially very useful in explaining the strong magnetic fields observed in our Galactic center, in starburst galaxies and in protogalactic environments. That is especially so because the centers of disk galaxies do not have strong shear flows. In light of that fact, one is inclined to ask whether a critical SN rate that is somewhere between the range of 12 to 40 times the galactic rate should be viewed as a hard limit for magnetic field amplification? There are many astrophysical effects that suggest otherwise. First, the galactic center as well as the centers of starbursts and protogalaxies have considerably denser gas that might be able to contain the remnants more efficiently without undergoing thermal runaway. Second, the stratification introduces preferred channels along which hot gas may leave the system. The existence of such preferred channels might also help dynamo action because it permits small scale tangled magnetic fields to leave the system as suggested by Blackman & Field (2000) and Kulsrud (2000). Third, fuelling by bars might also dredge in new molecular gas which might help keep the mean temperature low. Thus a SN rate that is twelve times the galactic SN rate should not be viewed as a hard bound. Even starburst systems like Arp 220 which have very high rates of star formation in their centers might be able to draw on this mechanism. A similar result could hold true for protogalaxies.

KA point out that the fundamental problem with applying mean field dynamo theory to our Galaxy stems from the fact that our Galaxy is a system with a very high Reynolds number. They argue that the high Reynolds number causes the small-scale magnetic energy in the turbulence to quench the small-scale helical flows in the turbulent velocity field. Their argument remains valid for any weakly driven turbulence. Our SN-driven turbulence simulations show that the Galaxy might be able to bypass the limitation because SN-driven shocks are strong at small scales, and so may not be subject to such $\alpha$-quenching arguments.

Recent EGRET gamma-ray observations show that it is difficult to find a source for the strong magnetic fields observed in the intracluster medium (ICM) of various clusters (Dolag, Bartelmann, & Lesch 1999). However, Kim, Kronberg & Tribble (1991) have made observations of the magnetic field in clusters. It might be pointed out that galaxies in a cluster are strongly susceptible to tidal interactions which is a strong source of central fuelling and starburst activity. Prolific and episodic starburst activity in the ICM can also amplify the magnetic field in the ICM. The superwinds have temperatures of $10^8$ K, (Suchkov et al. 1994). When these superwinds plow through the ICM, which has a much lower temperature of $10^{6-7}$ K, they have an effect that is similar to the interaction of SN shells with the ISM. Thus a mechanism like the one proposed here may also have applicability to the generation of magnetic fields in the ICM. It is also worth pointing out that Balsara, Livio and O'Dea (1994) have also studied the turbulent wakes of galaxies that propagate through the ICM. They found that the turbulent wakes are indeed strong sources of vorticity and helicity. Thus there are several mechanisms available for sustaining turbulence in the ICM.

Molecular clouds probably contain supersonic, trans-Alfvenic turbulence, see Crutcher (1999), indicating that the magnetic fields are roughly close to their equipartition values. Kornreich and Scalo (2000), Mac Low & Klessen (2004), Elmegreen and Scalo (2004) and Scalo and Elmegreen (2004) mention that strong shocks associated with high Mach number winds and jets in molecular clouds might play a role in sustaining the turbulence in molecular clouds. Molecular clouds are made clumpy by turbulent compression and gravitational collapse of protostellar cores (Mac Low 1999; Balsara, Ward-Thompson & Crutcher 2001). Thus the interaction of strong shocks with the clumpy molecular cloud material could operate in a fashion that is analogous to SN shocks propagating through the clumpy ISM. It is, therefore, possible that shock-driven turbulence in molecular clouds plays an important role in field amplification there.

**6.b) Conclusions**

This work provides several new insights into the problem of magnetic field amplification in a strongly turbulent environment. They are:

1) The SN-driven, turbulent, multiphase ISM possesses one of the major features needed for the operation of a fast dynamo: a robust and persistent helicity-generation mechanism. This implies that there are many interesting astrophysical systems where it is possible to produce such flows.

2) We confirm the report of BBC that strong SN shocks interacting with background interstellar turbulence produce large kinetic helicity fluctuations about a zero mean. In our model the turbulence is self-consistently generated by the SNs rather than having been generated by a uniform driver. The helicity fluctuations come about because of the amplification of turbulence in a strong shock and also because shocks propagating through a clumpy and turbulent medium naturally generate vorticity and helicity.

3) Seed magnetic fields amplify rapidly in such environments. The magnetic energy grows with growth times that are a small multiple of the eddy turnover time in the simulated system. In the linear phase of growth, different Fourier modes of the magnetic field with length scales both larger and smaller than the effective forcing scale grow at the same rate.

4) The quasi-linear phase of evolution (as defined by ZPP), where the initially fast linear growth of field is slowed down, sets in when the magnetic energy is about 1% of the fluid kinetic energy. This suggests that most of the growth in the protogalactic magnetic energy takes place rapidly during the linear phase of growth.

5) Spectral analysis of the flow field reveals that the solenoidal part of the total kinetic energy exceeds the compressible part of the total kinetic energy by almost an order of magnitude at most scales. It is remarkable that even though the ISM is kept turbulent by strongly compressible motions, the resultant supersonic turbulence has so much kinetic energy in solenoidal motions.

6) Our results may help explain why damped Ly-$\alpha$ systems have strong observed magnetic fields, and how protogalaxies can rapidly amplify their seed magnetic fields. Rapid magnetic field generation in molecular clouds, the galactic center, starburst galaxies and even in intracluster gas can be explained using the results discussed here.

**Acknowledgements** : DSB acknowledges support via NSF grants R36643-7390002, AST-005569-001 and DMS-0204640. JK was also supported by KOSEF through Astrophysical Research Center for the Structure and Evolution of the Cosmos (ARCSEC). M-MML acknowledges support from NSF grants AST-99-85392 and AST-03-07854. GJM is supported by DOE grant DE-FG02-94ER-40823. The majority of simulations were performed on PC clusters at UND and KAO but a few initial simulations were also performed at NCSA. The Korean cluster was acquired with funding from KAO and ARCSEC. The authors wish to thank the anonymous referee for several helpful comments.

# References

Balsara, D.S. 2000, Rev. Mex. Astron. Astrof., 9, 92

Balsara, D.S. 2004, ApJ Supp., 151, 149

Balsara, D.S., Benjamin, R. & Cox, D.P. 2001, ApJ, 563, 800 (BBC)

Balsara, D.S. & Kim, J.S. 2004 ApJ, 602, 1079

Balsara, D.S., Kim, J.S. and Mathews, G.J. 2004, "Amplification of Magnetic Flux and its Relation to Dynamical Chaos in Supernova-Driven Turbulence", in preparation.

Balsara, D.S., Ward-Thompson, D. and Crutcher, R.M. 2001, MNRAS, 227, 715

Balsara, D.S., Livio, M. & O'Dea, C.P. 1994, ApJ, 437, 83

Balsara, D.S. & Pouquet, A. 1999, Phys. Plasmas, 6, 89

Balsara, D.S., & Spicer, D.S. 1999a, J. Comp. Phys., 148, 133

Balsara, D.S., & Spicer, D.S. 1999b, J. Comp. Phys., 149, 270

Beck, R., Brandenburg, A., Moss, D., Shukurov, A. & Sokoloff, D. 1996, Ann. Rev. Astron. Astroph., 34, 155

Biermann, L. 1950, Z. Naturforsch., 5a, 65

Boldyrev, S., 2002, ApJ, 569, 841

Childress, S. & Gilbert, A. 1995, Stretch, Twist, Fold: The Fast Dynamo (Springer-Verlag, Berlin)

Churchill, C. W. & Le Brun, V. 1998, ApJ, 499, 677

Cox, D.P. & Smith, B.W. 1974, ApJ, 189, L105

Crutcher, R.M. 1999, ApJ, 520, 706

Dolag, K., Bartelmann, M. & Lesch, H. 1999, A&A, 348, 351

Elmegreen, B.G. and Scalo, J., 2004, "Interstellar Turbulence I: Observations and Processes", Ann. Rev. Astron. Astroph., to be published (astro-ph/0404451)

Ferrière, K. 1993a, ApJ, 404, 162

Ferrière, K. 1993b, ApJ, 409, 248

Ferrière, K. 1995, ApJ, 441, 281

Ferrière, K. 1998, A&A, 335, 488

Ferrière, K. & Schmitt, D. 2000, A&A, 358, 125

Field, G.B., Goldsmith, D.W. & Habing, H.J. 1969, ApJ, 155, L149

Fosalba, P., Lazarian, A., Prunet, S. and Tauber, J.A., 2002, ApJ, 564, 762

Gaensler, B.M. et al. 2001, ApJ, 549, 959

Galloway, D.J., and Procter, M.R.E. 1992, Nature, 356, 691

Han, J.L., Ferriere, K. & Manchester, R.N., 2004, 610, to appear (astro-ph/0404221)

Haverkorn, M., Katgert, P. & de Bruyn, A.G. 2000, A&A, 356, L13

Howard, A.M. & Kulsrud, R.M. 1997, ApJ, 483, 648

Kim, J.S., Balsara, D.S. & Mac Low, M.-M. 2001, J. Korean Ast. Soc., 34, 333

Kim, K.-T., Kronberg, P.P. & Tribble, P.C. 1991, ApJ, 379, 80

Kornreich, P. & Scalo, J. 2000, ApJ, 531, 366

Korpi, M.J. et al. 1999, ApJ, 514, L99

Kraichnan, R. & Nagarajan, S. 1967, Phys. Fluids, 10, 859

Kronberg, P.P. & Perry, J.J. 1982, ApJ, 263, 518

Kronberg, P.P. 1995, Nature, 374, 404

Kulsrud, R.M. & Anderson, S.W. 1992, ApJ, 396, 606 (KA)

Kulsrud, R.M., Cen, R., Ostriker, J.P. & Ryu, D. 1997, 480, 481

Kulsrud, R.M. 2000, "A Critical Review of Galactic Dynamos", Ann. Rev. A.&A.

Ledoux, C., Srianand, R. and Petitjean, P., 2002, Astron. & Astroph., 392, 781L

Mac Low, M.-M. & McCray, R. 1988, ApJ, 324, 776

Mac Low, M.-M. 1999, ApJ, 524, 169

Mac Low, M.-M., Balsara, D.S., Avillez, M. & Kim, J.S. 2004, "The Distribution of Pressures in a SN-Driven Interstellar Medium", ApJ, submitted (astro-ph/0106509)

Mac Low, M.-M, & Klessen, R. S. 2004, Rev. Mod. Phys., 76, 125

Magnier, E. A., Chu, Y.-H., Points, S. D., Hwang, U., & Smith, R. C. 1996, ApJ, 464, 829

Maron, J. & Blackman, E.G. 2002, ApJ, 566, L41

McCray, R. & Kafatos, M. 1987, ApJ, 317, 190

McKee, C.F. & Ostriker, J.P. 1977, ApJ, 218, 149

McKee, C.F. & Zweibel, E.G. 1995, ApJ, 440, 686

McKenzie, J.F. & Westphal, K.O. 1968 Phys. Fluids, 11, 2350

Minter, A.H. & Spangler, S.R. 1996, ApJ, 458, 194

Mushotzky, R. F. and Loewenstein, M. 1997, Ap.J.Lett., 481, L63

Parker, E. N. 1971, ApJ, 163, 255

Perry, J.J., Watson, A. M., & Kronberg, P.P. 1993, ApJ, 406, 407

Passot, T. and Pouquet, A. 1987, J Fluid Mech., 181, 441

Porter, D.H., Pouquet, A., Sytine, I. & Woodward, P.R. 1999, Physica A, 263, 263

Pouquet, A., Frisch, U., & Léorat, J. 1976, J. Fluid Mech., 77, 321

Rand, R.J. & Kulkarni, S.R. 1989, ApJ, 343, 760

Rand, R.J., Kulkarni, S.R., & Hester, J. J. 1991, ApJ (Letters), 352, L1

Raymond, J.C. & Smith, B.W. 1977, ApJ Supp., 35, 419

Ruzmaikin, A*(*a)., Shukurov, A.M. & Sokoloff, D.D. 1988, "Magnetic Fields in Galaxies", (Kluwer Academic Publishers, Dordrecht)

Samtaney, R. & Zabusky, N.J. 1994, J. Fluid Mech., 269, 45

Scalo, J., and Elmegreen, B.G., 2004, "Interstellar Turbulence II: Implications and Effects", Ann. Rev. Astron. Astroph., to be published (astro-ph/0404452)

Spangler, S. 1999, in "Interstellar Turbulence" eds. Franco, J. & Carraminana, A., (Cambridge Univ. Press, Cambridge), 41

Steenbeck, M., Krause, F. & Radler, G. 1966, Z. Nat., A21, 369

Suchkov, A., Balsara, D.S., Heckman, T. and Leitherer, C. 1994, ApJ, 430, 511

Vainshtein, S.I., & Zeldovich, Ya.B. 1972, Sov. Phys. Usp., 15, 159

Vestuto, J.G., Ostriker, E.C. & Stone, J.M. 2003, ApJ, 590, 857

Vishniac, E.T. 1983, ApJ, 274, 152

Williams, R. M., Chu, Y.-H., Dickel, J. Smith, R. C., Milne, D. K., & Winkler, P. F. 1999, ApJ, 514, 798

Wolfe, A.M., Lanzetta, K.M. & Oren, A.L. 1992, ApJ, 388, 17

Wolfire, M.G., McKee, C.F., Hollenbach, D. & Tielens, A.G.G.M. 1995, ApJ, 443, 152

Wolfire, M.G., McKee, C.F., Hollenbach, D. & Tielens, A.G.G.M. 2003, ApJ, 587, 278

Zeinecke, E., Politano, H., & Pouquet, A. 1998, Phys. Rev. Lett., 81, 4640 (ZPP)

Figure Captions

1. Log of magnetic energy (*solid*) and rms density (*dotted*) versus time for runs at $128^3$ and $256^3$ zone resolution. Thick and thin lines indicate high and low resolution simulations respectively.

2. Growth time smoothed with 10 Myr window function for the runs in Figure 1.

3. Growth of magnetic energy in the $256^3$ zone simulation on all scales smaller (*dotted*) and larger (*solid*) than the typical size of a SN remnant, 30 pc, equivalent to $k/k_0 = 7$.

4. Time evolution of spectral modes with wavenumber $k/k_0 = 3$ (*solid*), 5 (*dotted*), 9 (*dashed*) and 11 (*long dashed*). The first two modes represent length scales larger than the size of a SN remnant (roughly $k/k_0 = 7$), while the second two are smaller.

5. Evolution of rms fluctuations in the kinetic (*dotted*) and magnetic (*solid*) helicities as a function of time.

6. Time evolution of thermal (green), kinetic (blue) and total (red) energy in runs with SN rates of *(a)* eight, *(b)* twelve, and *(c)* forty times the present Galactic value, as well as *(d)* magnetic energy in the last of these runs.

7. Shows the spectra for the density-averaged kinetic (*solid*) and magnetic (*short dashed*) energies and fluid density (*long dashed*) at several different times in the $256^3$ zone run. Different epochs are shown with different colors. We see that the kinetic energy and fluid density saturate while the magnetic energy keeps growing.

8. Shows the spectrum for the compressible (*solid*) and solenoidal (*dotted*) parts of the velocity field at various times from the $256^3$ zone run. Different epochs are shown with different colors. We find the surprising result that except for some of the largest length scales, the solenoidal part of the velocity spectrum overwhelms the compressible part of the velocity spectrum by more than an order of magnitude.

9. Absolute value of *(a)* magnetic and *(b)* kinetic helicity spectrum at various times in the simulation. Despite their jaggedness, the magnetic helicity spectra show a secular increase with increasing time, consistent with the growth of magnetic energy. The kinetic helicity shows no such secular increase.

10. Images of the log of thermal (*upper left*) and magnetic (*upper right*) pressure, of the signed log of kinetic helicity sign($H_v$) log (|$H_v(x,y)$|) (*lower left*), and of magnetic helicity (*lower right*) in a slice plane that passes through the recent remnant at the bottom of the image. Note that the magnetic helicity is not logarithmically scaled.

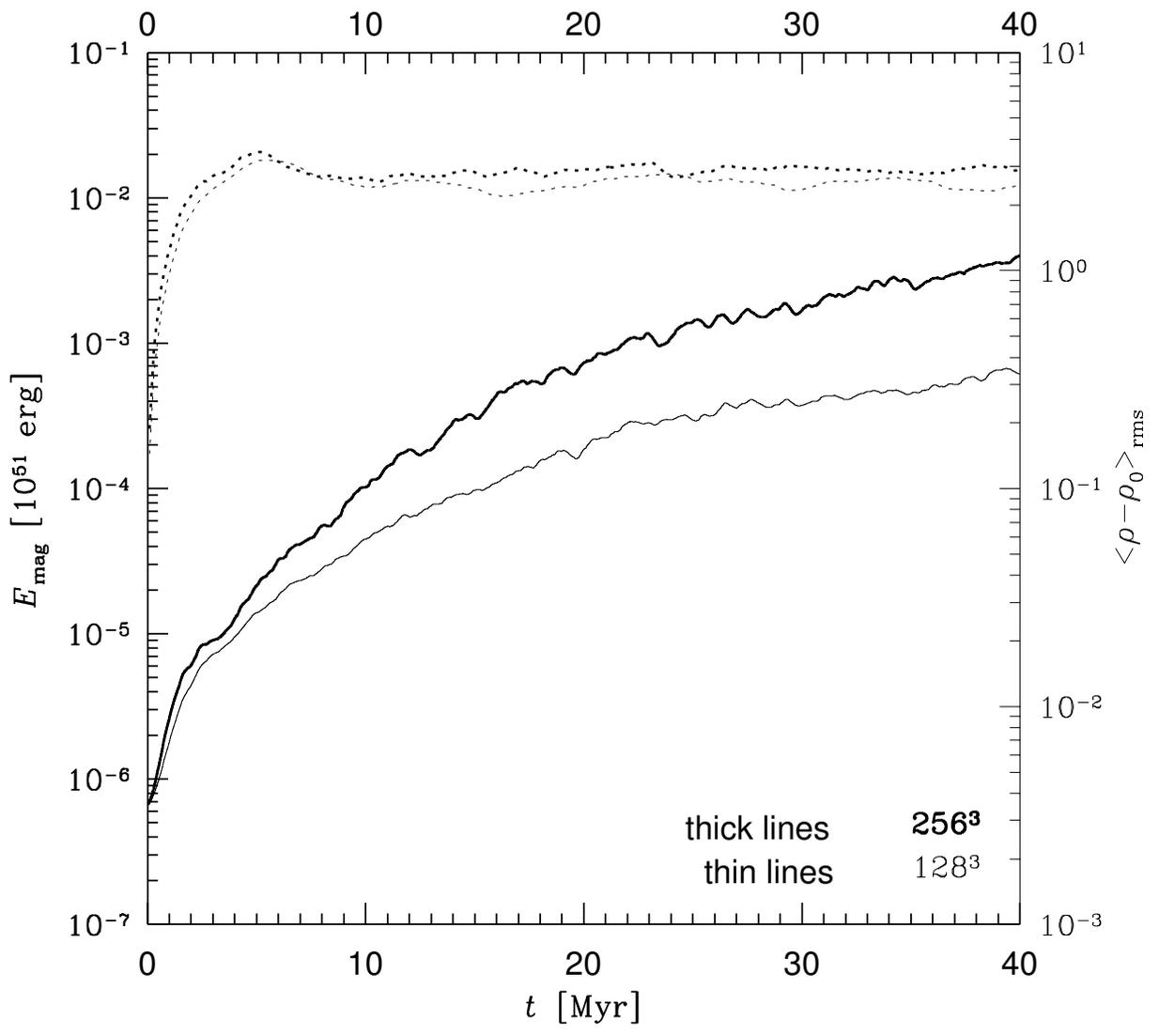

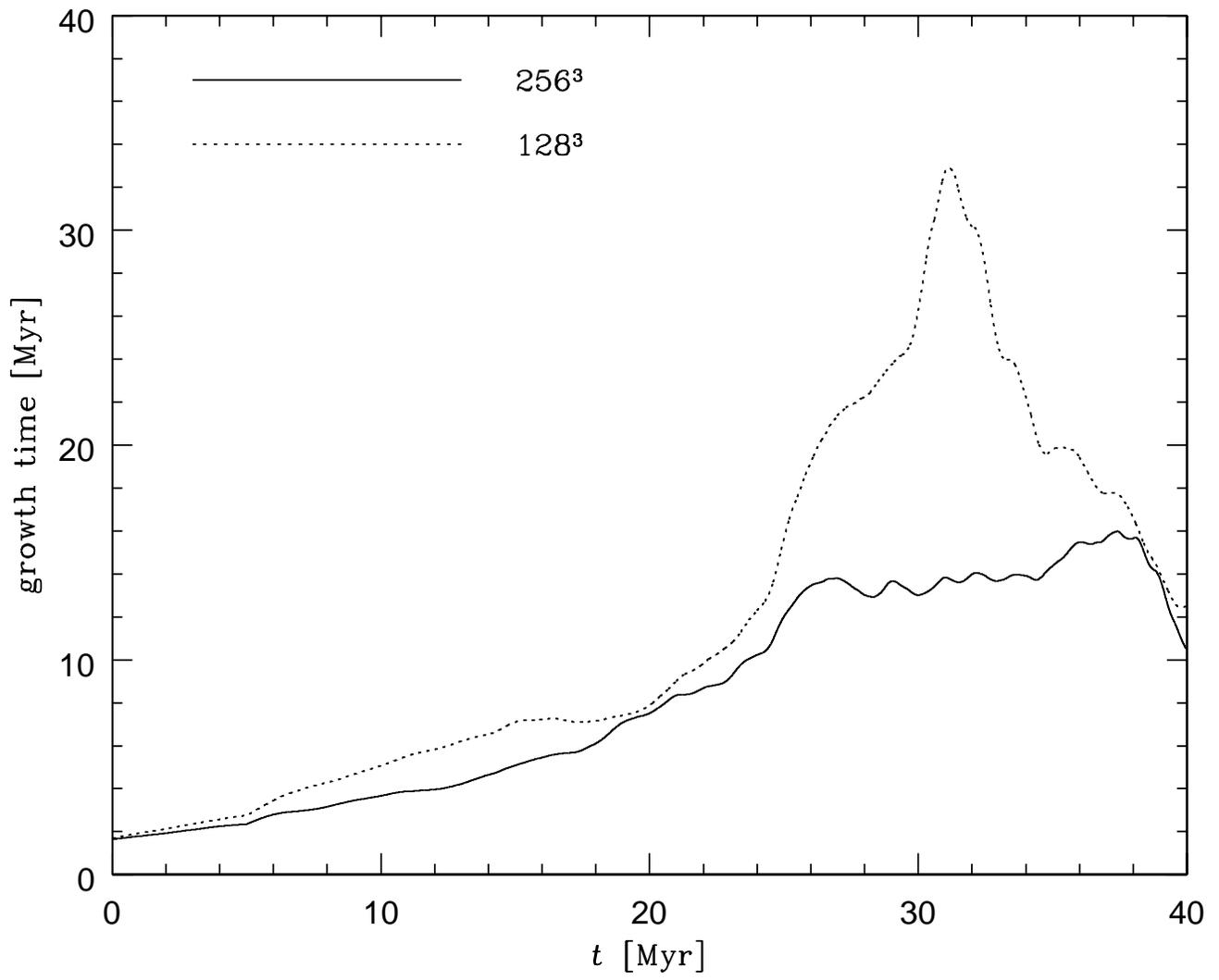

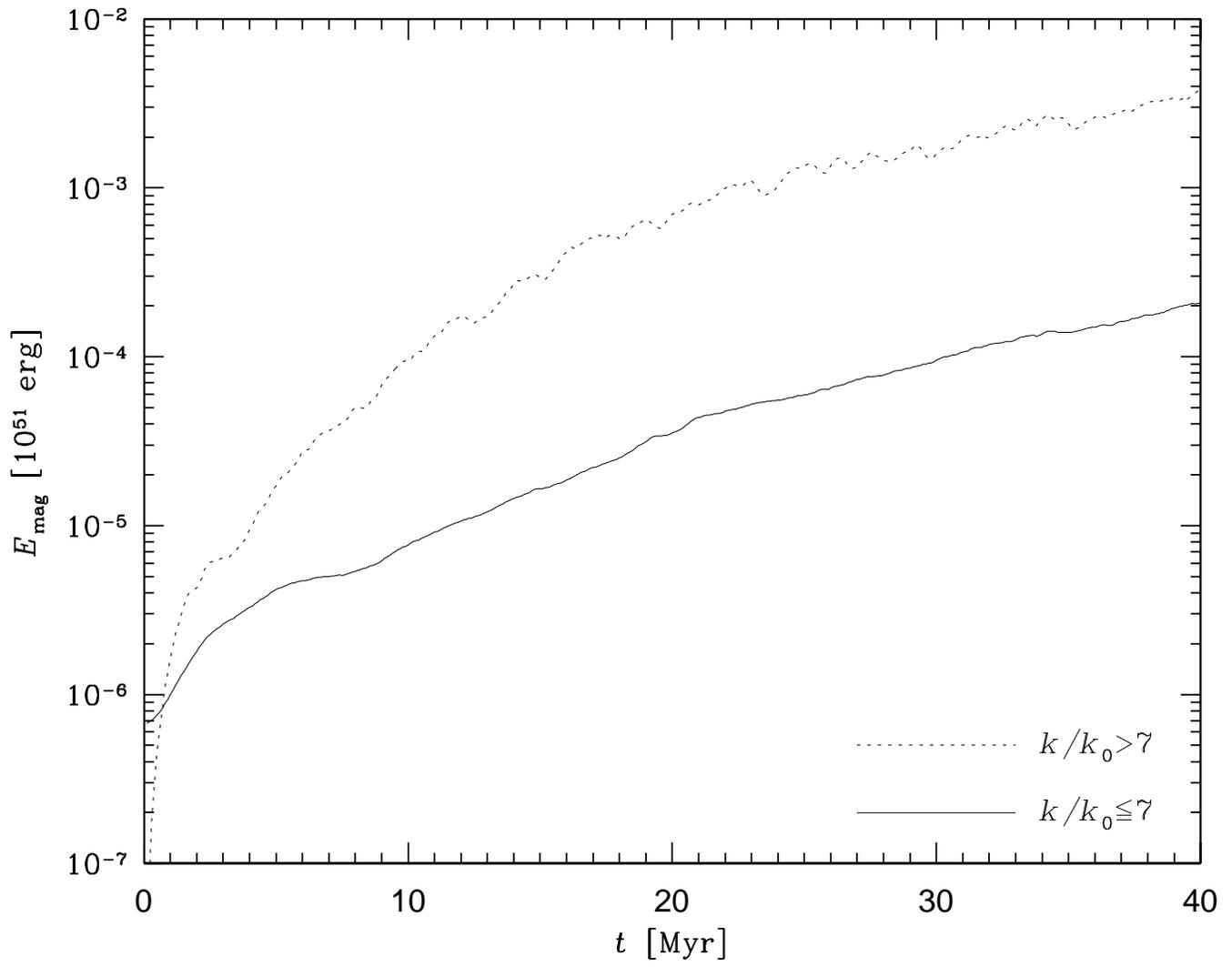

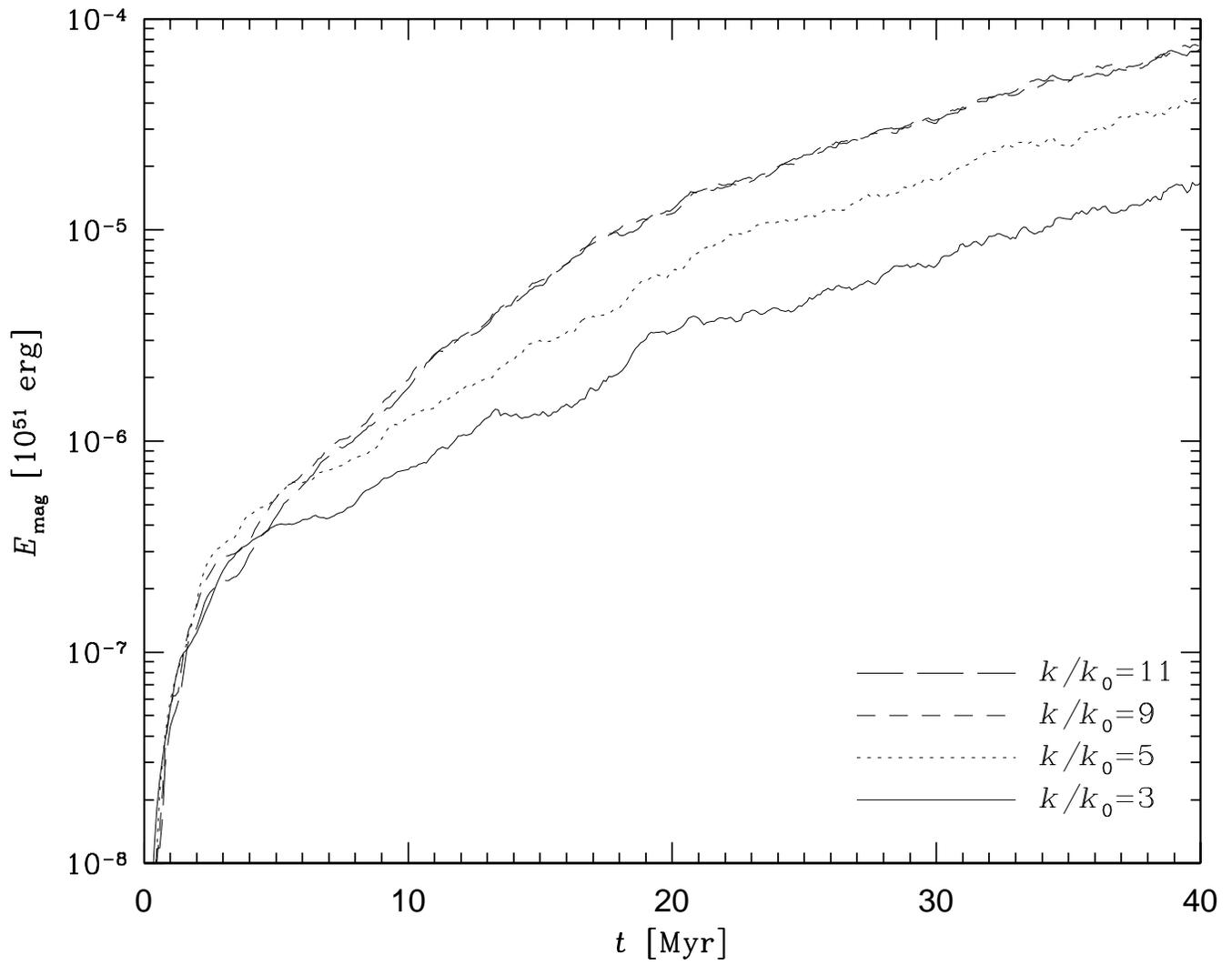

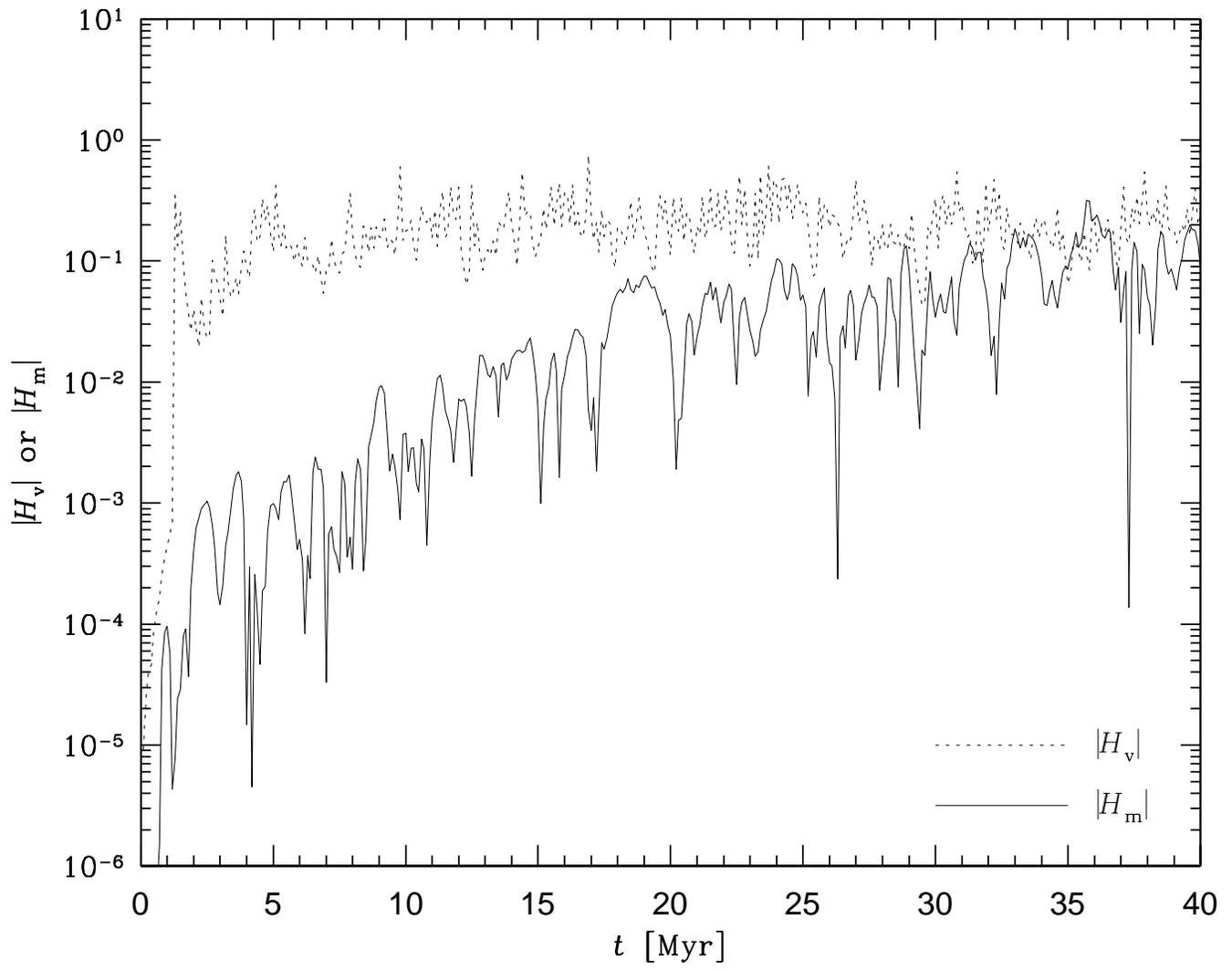

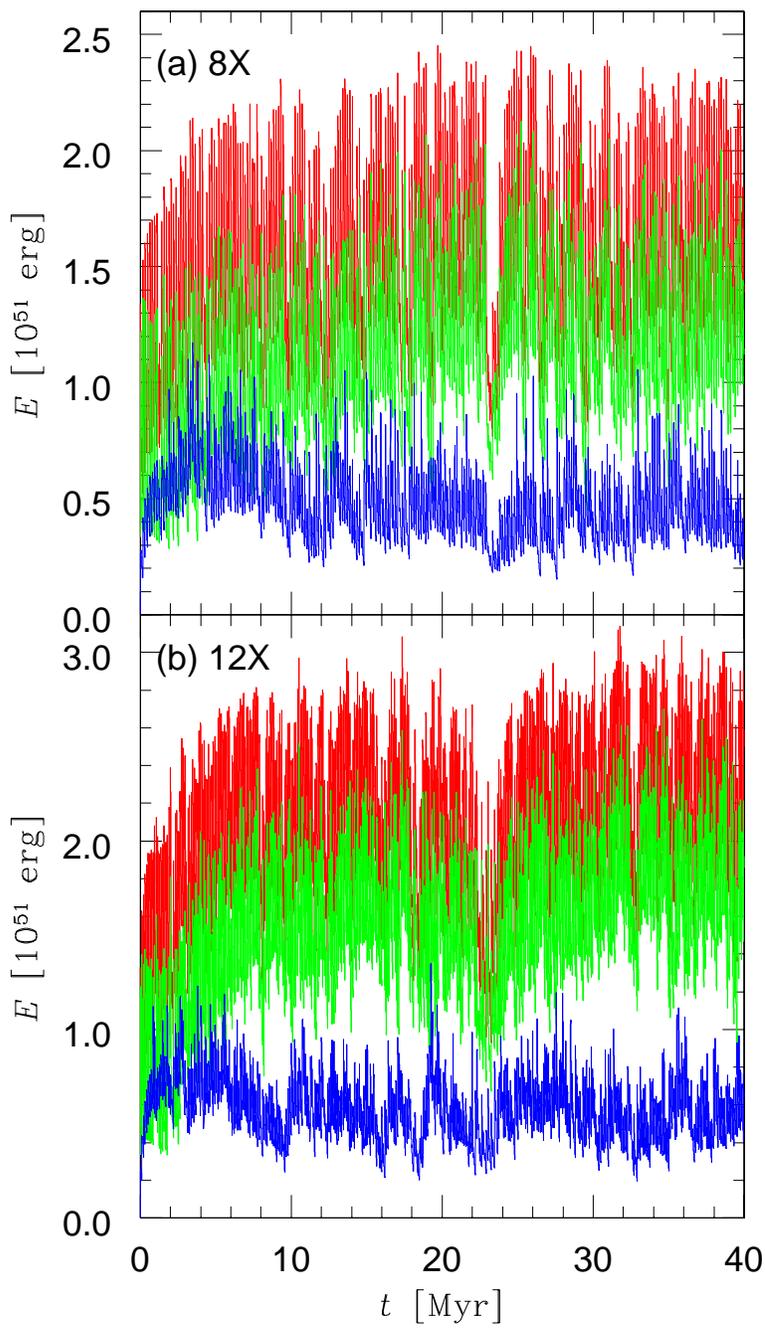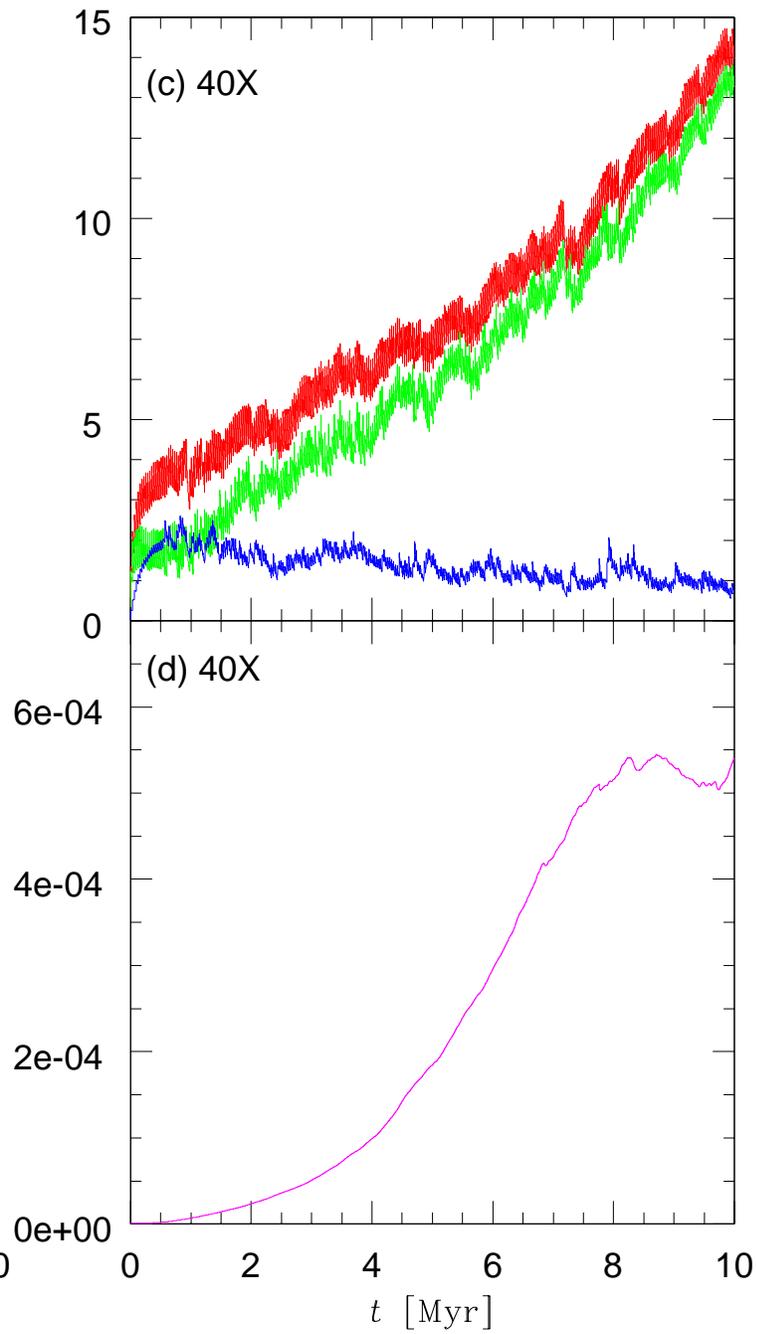

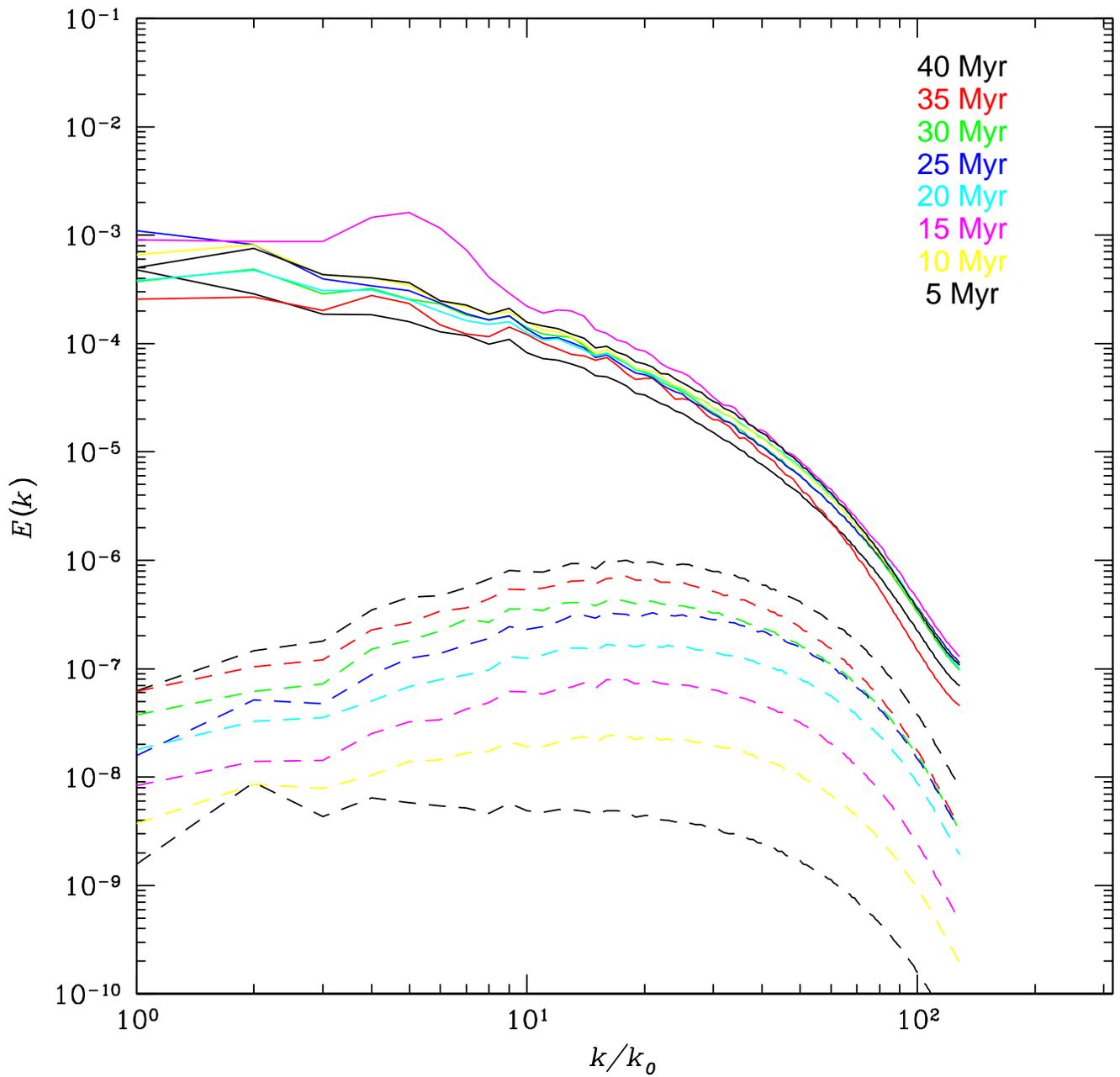

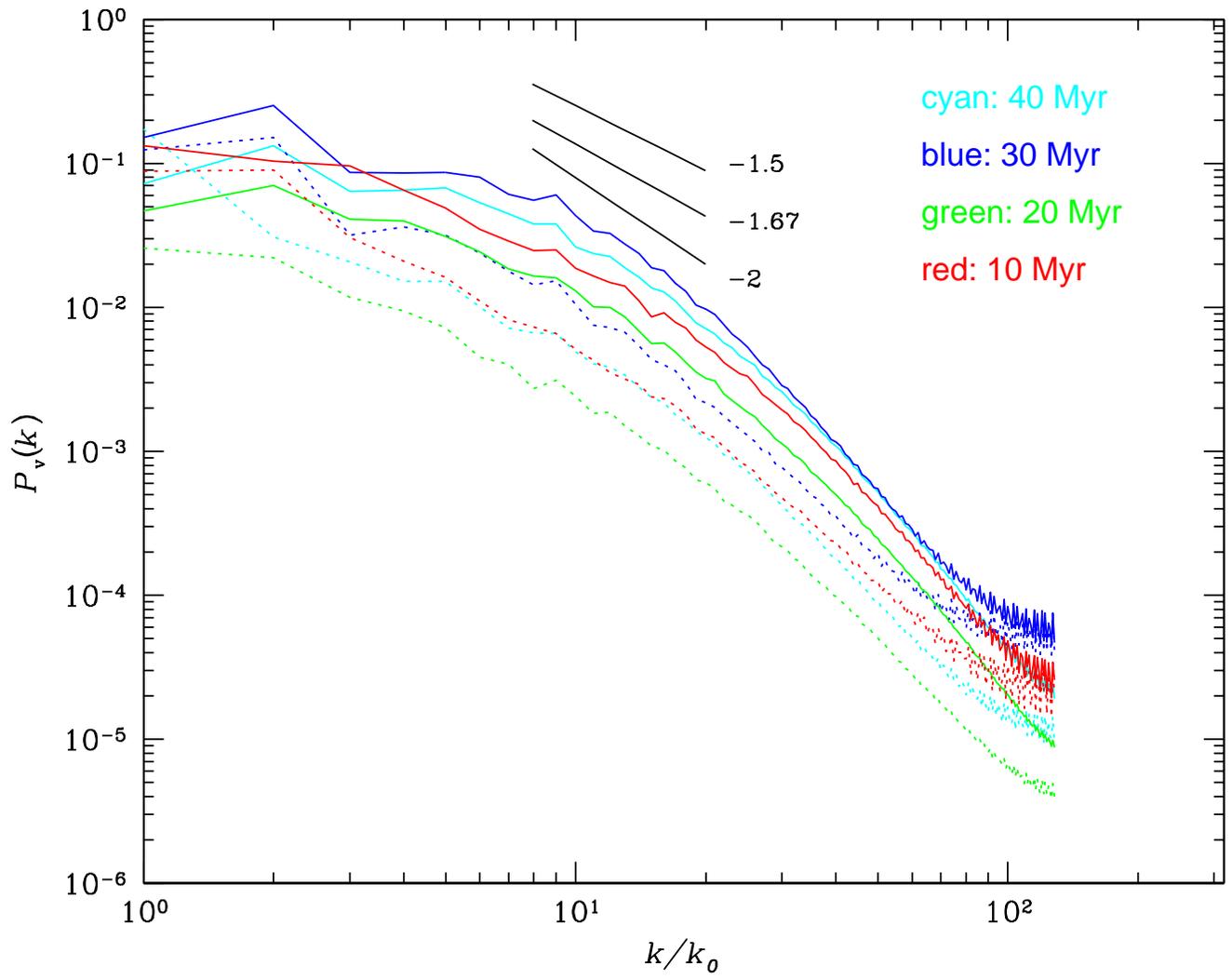

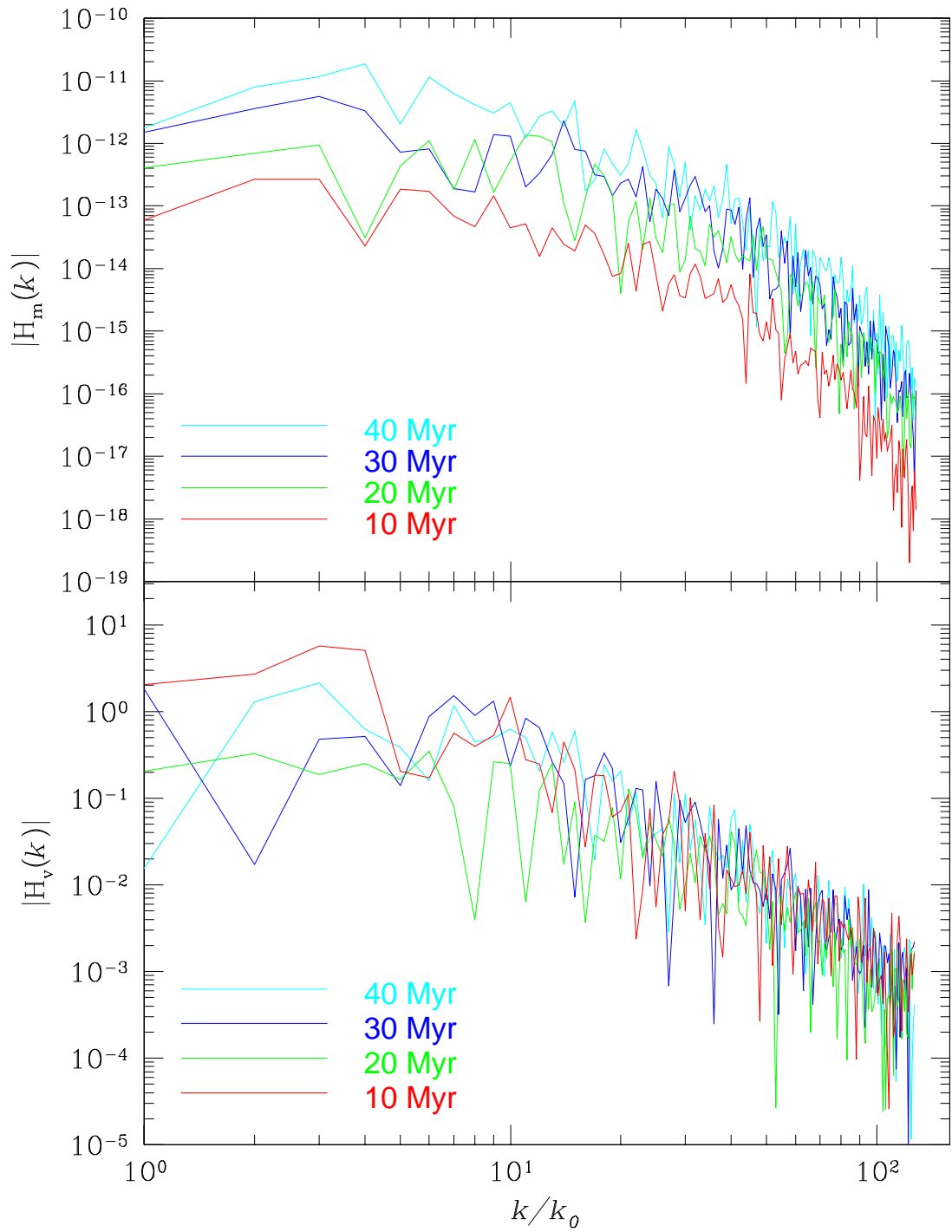

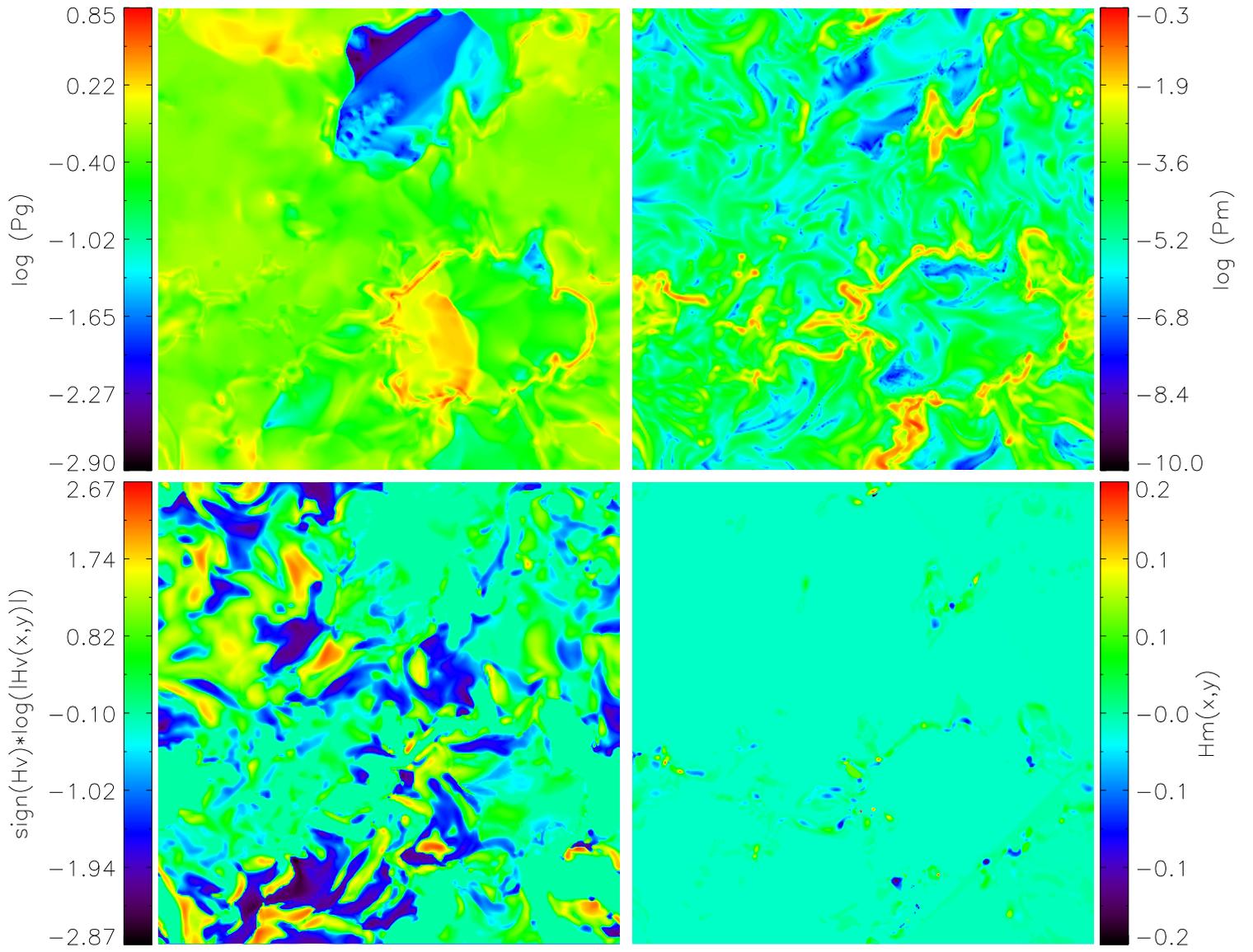